\documentclass[letterpaper]{article} 
\usepackage{aaai2026}  
\usepackage{times}  
\usepackage{helvet}  
\usepackage{courier}  
\usepackage[hyphens]{url}  
\usepackage{graphicx} 
\usepackage{natbib}  
\usepackage{caption} 
\usepackage{algorithm}
\usepackage{algorithmic}
\usepackage{amsmath}
\usepackage{amssymb}
\usepackage{amsthm}
\usepackage{booktabs}
\usepackage{enumitem}
\usepackage{graphicx}
\usepackage{color}
\usepackage{times}
\usepackage{multirow}
\usepackage{subfig}
\usepackage{booktabs}

\usepackage{newfloat}
\usepackage{listings}
\DeclareCaptionStyle{ruled}{labelfont=normalfont,labelsep=colon,strut=off} 
\floatstyle{ruled}
\newfloat{listing}{tb}{lst}{}
\floatname{listing}{Listing}
\title{Multi-Aspect Cross-modal Quantization for Generative Recommendation}
\author{
    Fuwei Zhang$^{1}$,
    Xiaoyu Liu$^{1}$,
    Dongbo Xi$^{2}$,
    Jishen Yin$^{2}$,
    Huan Chen$^{2}$,
    Peng Yan$^{2}$,
    Fuzhen Zhuang$^{1,3}$,
    Zhao Zhang$^{3}$\thanks{Corresponding author}
}
\affiliations{
    \textsuperscript{\rm 1}Institute of Artificial Intelligence, Beihang University\\
    \textsuperscript{\rm 2}Meituan\\
    \textsuperscript{\rm 3}SKLCCSE, School of Computer Science and Engineering, Beihang University

    \{zhangfuwei, liuxiaoyv, zhuangfuzhen\}@buaa.edu.cn, \{xidongbo,yinjishen,chenhuan15,yanpeng04\}@meituan.com, zhangzhao.cs.ai@gmail.com
}

\usepackage{bibentry}

\begin{document}

\maketitle

\begin{abstract}
Generative Recommendation (GR) has emerged as a new paradigm in recommender systems. This approach relies on quantized representations to discretize item features, modeling users’ historical interactions as sequences of discrete tokens. Based on these tokenized sequences, GR predicts the next item by employing next-token prediction methods. The challenges of GR lie in constructing high-quality semantic identifiers (IDs) that are hierarchically organized, minimally conflicting, and conducive to effective generative model training. However, current approaches remain limited in their ability to harness multimodal information and to capture the deep and intricate interactions among diverse modalities, both of which are essential for learning high-quality semantic IDs and for effectively training GR models. To address this, we propose \textbf{M}ulti-\textbf{A}spect \textbf{C}ross-modal quantization for generative \textbf{Rec}ommendation~(MACRec), which introduces multimodal information and incorporates it into both semantic ID learning and generative model training from different aspects. Specifically, we first introduce cross-modal quantization during the ID learning process, which effectively reduces conflict rates and thus improves codebook usability through the complementary integration of multimodal information. In addition, to further enhance the generative ability of our GR model, we incorporate multi-aspect cross-modal alignments, including the implicit and explicit alignments. Finally, we conduct extensive experiments on three well-known recommendation datasets to demonstrate the effectiveness of our proposed method.
\end{abstract}

\begin{links}
    \link{Code}{https://github.com/zhangfw123/MACRec}
\end{links}


\section{Introduction}
Recommendation systems play a crucial role in helping users navigate information overload by providing personalized item suggestions. As a result, they have now been widely applied across various domains, such as e-commerce~\cite{amazonrec1,taobaorec2,liu2025multi,zhang2022mind,xi2020graph,zhang2024temporal,zhang2022latent,chen2024fairgap,chen2025fairdgcl}, social media platforms~\cite{davidson2010youtube,dnn}, online advertising~\cite{xi2019modelling,xi2021modeling}, and short video platforms~\cite{zhu2024interest,zhu2025long,bin2025real}. 

With the development of large language models (LLMs), numerous LLM-enhanced recommendation approaches~\cite{du2024enhancing} have been proposed. Among them, a new paradigm called generative recommendation (GR) has recently gained significant attention. In GR, the recommendation task is reformulated as a next-token prediction problem, where the model is provided with a sequence of item semantic identifiers representing a user's interaction history and is tasked with generating the semantic identifier of the next item that aligns with the user's interests. By harnessing the powerful natural language understanding and sequence modeling capabilities of LLMs, GR offers enhanced flexibility and expressiveness, enabling more accurate modeling of user intent and richer contextual integration~\cite{li2024large,zhang2025hiergr,zhang2025multi,liu2025cat}. 

\begin{figure}[t]
    \centering
    \subfloat[Text Embeddings\label{fig:left}]{
        \includegraphics[width=0.45\linewidth]{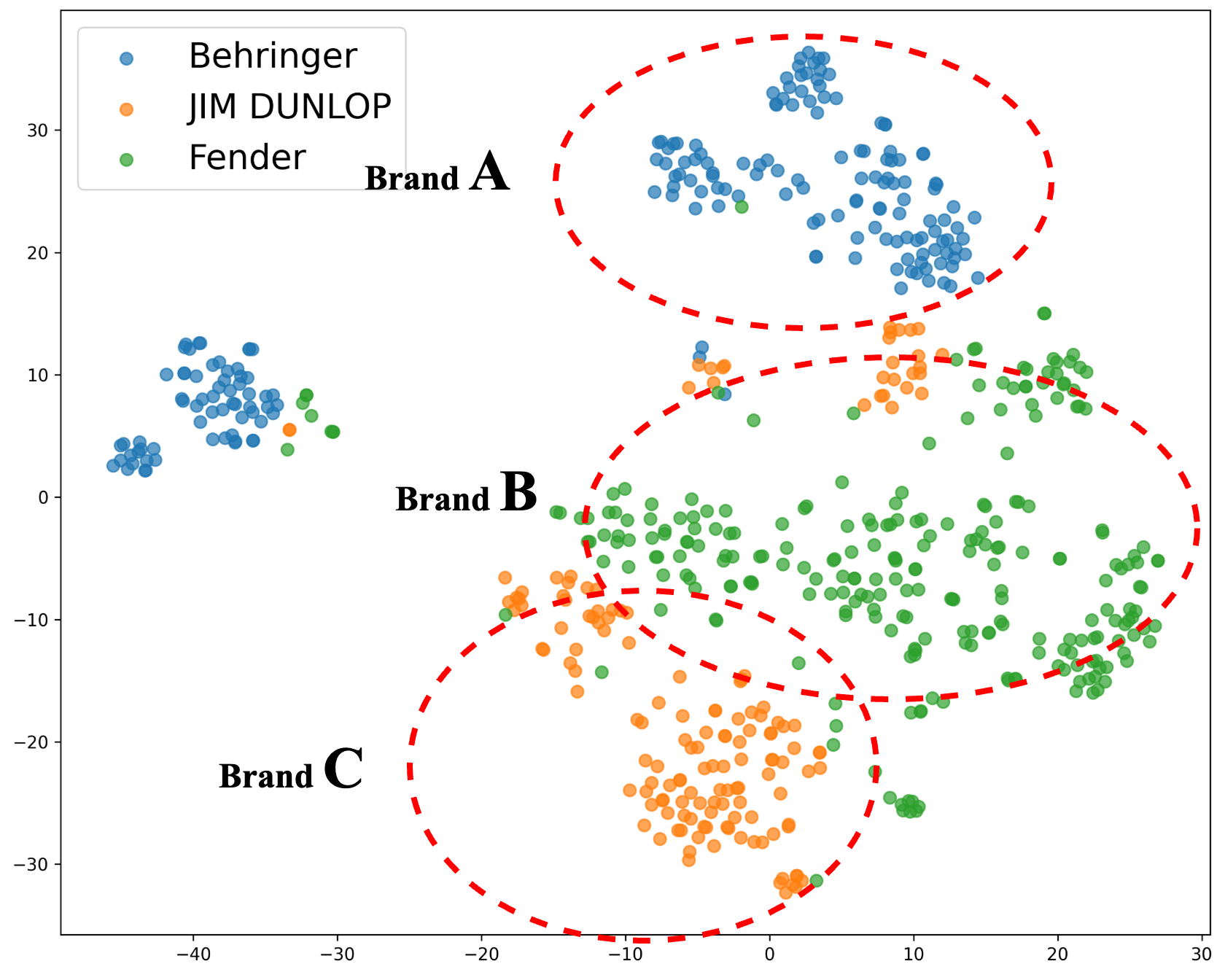}
    }
    \hfill
    \subfloat[Image Embeddings\label{fig:right}]{
        \includegraphics[width=0.45\linewidth]{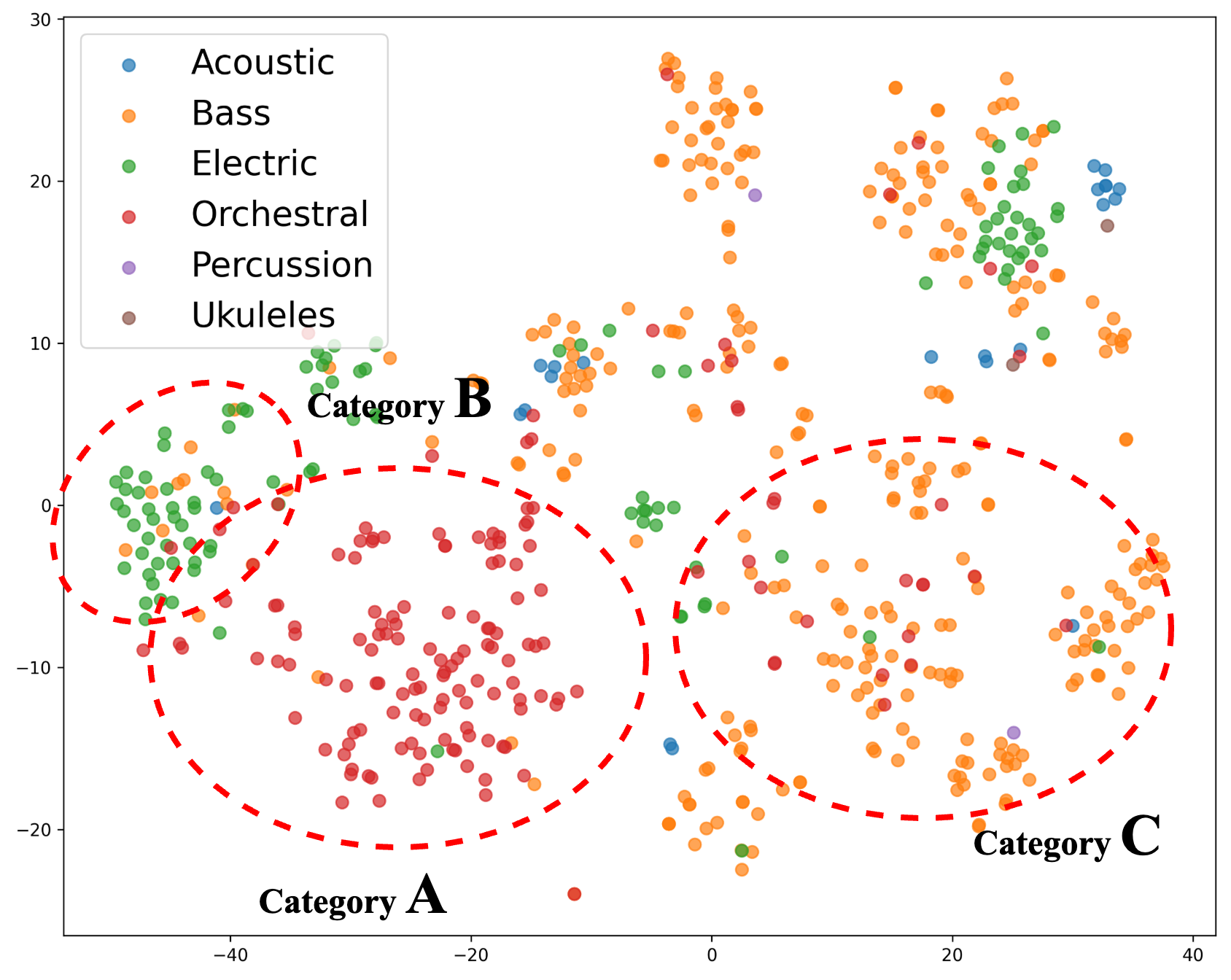}
    }
    \caption{Visualization examples of different modality embeddings on Instrument.}
    \label{fig:intro}
\end{figure}

A number of studies have explored the paradigm of GR. These works primarily focus on quantizing item embeddings into unique semantic identifiers~\cite{tiger,letter,mql}, and subsequently constructing user interaction sequences via the item semantic ID sequences. In this framework, recommendations are produced by generating the semantic IDs of recommended items. Existing approaches predominantly emphasize the discretization of textual embeddings to form these semantic IDs. However, relying solely on embeddings derived from a single modality might lead to limited semantic discriminability. For example, products from the same brand, such as different types of instruments produced by that brand, often have smaller distances between their encoded representations due to the prominence of brand-related textual features. This reduces the distinctiveness of each instrument, making it more difficult to differentiate between products of the same brand with different functions. 
Such interference may reduce the quality of semantic ID representations, potentially affecting recommendation performance.
Moreover, commonly used quantization methods such as Residual Quantization Variational Autoencoder (RQ-VAE) are prone to significant semantic loss in deeper hierarchical structures~\cite{zhou2025onerec,zhou2025onerecv2}. This semantic loss may cause the model to lack clear semantic guidance when assigning tokens, resulting in an almost random distribution and thereby weakening the semantic hierarchy of the generated item representations.

To address this challenge, we incorporate multimodal information, including images, to overcome the limitations of textual representations. Visual features in images, such as shape and color, are often more distinctive and sensitive than textual descriptions, thereby providing more effective support for understanding and conveying information.
In Figure~\ref{fig:intro}, we visualize the embeddings generated from text and images, respectively. The results indicate that text-based embeddings are more effective at clustering items from the same brand, while image-based embeddings excel at distinguishing between different types of musical instruments. These observations suggest that different modalities capture complementary aspects of user preferences. Based on these findings, we argue that relying solely on a single modality is insufficient to comprehensively represent the semantic characteristics of items. Thus, we integrate multimodal item information into the GR to enhance semantic expressiveness and improve the model’s generative capability.

To address these challenges, we propose \textbf{M}ulti-\textbf{A}spect \textbf{C}ross-modal quantization for generative \textbf{Rec}ommendation~(\textbf{MACRec}), a novel generative recommendation framework that effectively integrates multimodal information. By leveraging cross-modal information, MACRec learns hierarchically meaningful semantic IDs for items and enhances the training of GR models. 
Specifically, during the item quantization stage, we propose a cross-modal quantization method that incorporates cross-modal contrastive learning into each layer of residual quantization to enhance information interaction across different modalities and reduce semantic loss. Meanwhile, we leverage multimodal alignment to optimize the reconstructed representations, thereby further enhancing the representational capacity of the codebook. 
During the training phase of the GR model, we employ multi-aspect alignment strategies to enhance the model’s understanding of semantic IDs and to enable the learning of shared features across different modalities. These strategies include implicit alignment in the latent space through contrastive methods, as well as explicit alignment within the generative task. Through a series of cross-modal interactions, our model achieves significant improvements in recommendation performance.

Here, we summarize our contributions:
\begin{itemize}
    \item To capture richer and more discriminative semantics, we propose a novel cross-modal quantization method that integrates contrastive learning into residual quantization and reconstruction, yielding hierarchically meaningful semantic IDs for items.
    \item To enable the model to learn common features from different modalities, we employ multi-aspect alignment strategies, including both implicit alignment in the latent space and explicit alignment in the generative task. 
    \item We conduct experiments on three widely used recommendation datasets, and our approach significantly outperforms state-of-the-art GR models.
\end{itemize}


\section{Related Work}
In the related work section, we mainly introduce traditional sequential recommendation methods and generative recommendation methods under different modalities.

\subsection{Sequential Recommendation}

Single-modal sequential recommendation focuses on modeling user behavior sequences based solely on interaction data, aiming to capture users' dynamic preferences over time. Early approaches employ neural networks such as GRU4Rec~\cite{gru4rec}, STAMP~\cite{liu2018stamp}, and NARM~\cite{narm2017} to learn sequential patterns and temporal dependencies. The introduction of attention-based models like SASRec~\cite{sasrec} further improves the ability to model long-range dependencies within sequences. More recently, pretrained language models (PLMs) such as BERT4Rec~\cite{sun2019bert4rec} have significantly advanced performance by leveraging large-scale self-supervised pretraining. In addition, prompt-based methods like P5~\cite{p5} and M6-Rec~\cite{cui2022m6rec} reformulate recommendation tasks as language modeling problems, which further enhances model generalization and flexibility. 

Multi-modal sequential recommendation enriches sequential representations by incorporating various item modalities (e.g., text, images), thereby improving recommendation quality~\cite{liu2024multimodal}. Recent approaches commonly employ deep and graph neural networks, such as MMGCN~\cite{wei2019mmgcn} and GRCN~\cite{grcn}, to integrate heterogeneous features for enhanced user-item interaction modeling. Additionally, contrastive learning and multimodal pretraining methods such as MMGCL~\cite{yi2022multi} and MISSRec~\cite{wang2023missrec} further strengthen user interest modeling. The VIP5~\cite{vip5} framework extends prompt-based techniques to multimodal settings and has advanced the field’s performance.

\subsection{Generative Recommendation}
With the rapid development of large language models (LLMs), the potential of generative recommendation (GR) has attracted increasing attention. Early works such as TIGER~\cite{tiger} discretize item sequences into tokens, enabling generative recommendation paradigms. LC-Rec~\cite{LCRec} utilizes the natural language understanding abilities of LLMs to support diverse task-specific fine-tuning in recommendation.
LETTER~\cite{letter} extends TIGER by introducing collaborative filtering embeddings and an additional loss function to improve codebook utilization. In multimodal generative recommendation, MMGRec~\cite{liu2024mmgrec} uses a Graph RQ-VAE to generate item representations by integrating multimodal features with collaborative signals. MQL4GRec~\cite{mql} further advances the field by encoding multimodal and cross-domain item information into a unified quantized language, facilitating knowledge transfer and achieving better performance than previous methods.

However, existing multimodal GR models typically encode each modality separately to obtain semantic IDs for different modalities. They do not consider cross-modal interactions during the quantization process, which makes them more prone to hierarchical semantic loss. In this paper, we are the first to introduce cross-modal learning during quantization, enabling IDs to capture the advantageous features of different modalities. In addition, we also incorporate both implicit and explicit alignment methods during GR training, allowing features from different modalities to complement each other more effectively.


\begin{figure*}[t]
	  \centering
      \includegraphics[width=\linewidth]{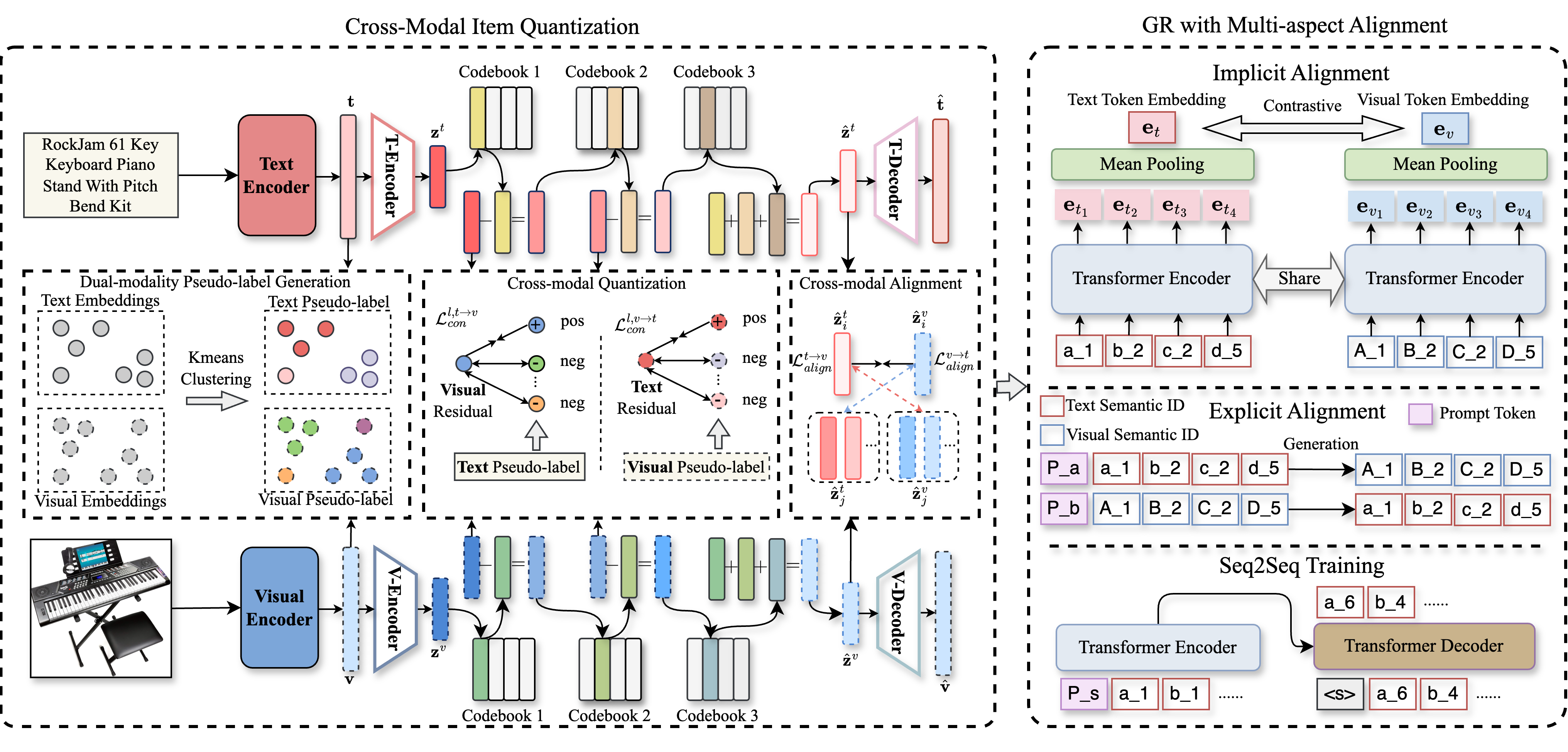}
      \caption{Overall architecture of MACRec. Left: Cross-modal Item Quantization, including Dual-modality Pseudo-label Generation, Cross-modal Quantization, and Cross-modal Reconstruction Alignment, aiming to generate high-quality semantic IDs across different modalities. Right: Generative Recommendation with Multi-aspect Alignment, aligning features from different modalities via implicit alignment, explicit alignment, and training the GR model with Seq2Seq task.}
      \label{fig:model}
\end{figure*}

\section{Methodology}
In this section, we organize our approach into two main modules. Figure~\ref{fig:model} illustrates the overall architecture, which includes a cross-modal item quantization module for generating discrete semantic IDs, and the training phase of the GR model with multi-aspect alignment.

\subsection{Cross-modal Item Quantization} 

Original RQ-VAE approaches primarily focus on quantizing a single embedding into discrete tokens. However, when dealing with multimodal information such as images and text, simply concatenating their embeddings and quantizing them as a unified token set, or independently quantizing each modality into separate tokens, both have some limitations. First, the dimensions of image and text embeddings often differ substantially, and straightforward concatenation tends to bias the quantization process toward the modality with higher dimensionality. Second, independently quantizing each modality and subsequently training the GR model fails to fully exploit the complementary nature of cross-modal information. In response to these challenges, we propose the Cross-modal Item Quantization framework. Our method effectively utilizes multimodal contrastive learning in the learning of quantization and reconstruction, enabling the generation of semantically discriminative item identifiers while simultaneously reducing identifier collision rates.  

\subsubsection{Dual-modality Pseudo-label Generation}
To perform quantization across different modalities using contrastive learning, we first generate pseudo-labels by clustering items according to their text and visual embeddings, which are then used to construct positive samples.
Specifically, for a given item $i$, its text information is encoded into an embedding $\mathbf{t}_i$ using an open-source large language model such as LLaMA~\cite{touvron2023llama}. Simultaneously, the visual content of the item's image is encoded into an embedding $\mathbf{v}_i$ using a Vision Transformer (ViT)~\cite{vit}.

Subsequently, we perform K-means clustering independently on the textual and visual embeddings, partitioning them into $K$ clusters. The clustering process can be formulated as follows:

\begin{equation}
    \mathcal{C}_\text{text} = \text{KMeans}(\{\mathbf{t}_i\}_{i=1}^N), \quad
    \mathcal{C}_\text{vision} = \text{KMeans}(\{\mathbf{v}_i\}_{i=1}^N),
\end{equation}
where $\mathcal{C}_\text{text}$ and $\mathcal{C}_\text{vision}$ denote the resulting cluster assignments (pseudo-labels) for the text and vision modalities, respectively, and $N$ is the total number of items.

\subsubsection{Cross-modal Quantization with Contrastive Learning}

Residual-Quantized Variational AutoEncoder (RQ-VAE)~\cite{image_rqvae,tiger} is an embedding quantization method that builds upon multi-layer vector quantization (VQ). In this framework, VQ discretizes continuous embeddings by mapping them to the closest entries in a learnable codebook, thereby effectively compressing the representation space. Residual quantization~(RQ) further enhances this process by sequentially applying multiple VQ layers, where each layer performs VQ on the residuals from the previous layer.
In multimodal scenarios, it is necessary to quantize information from different modalities. For a given item, both the text and visual embeddings are first processed by an encoder composed of a multi-layer perceptron (MLP) to obtain latent representations, which are denoted as $\mathbf{z}^t = \text{T-Encoder}(\mathbf{t})$ and $\mathbf{z}^v = \text{V-Encoder}(\mathbf{v})$, respectively. 
Here, $\mathbf{z}^t$ and $\mathbf{z}^v$ represent the latent representation for text and visual information, respectively, which are used as the residuals ($\mathbf{r}^t_0 = \mathbf{z}^t, \mathbf{r}^v_0 = \mathbf{z}^v$) for quantization in the first VQ layer. 

At the $l$-th layer, there is a learnable codebook $C_l^{v/t} = \{\mathbf{e}_{l,k}^{v/t}\}_{k=0}^{M}$ for each modality, where $M$ denotes the codebook size. For each modality, the residual at the $l$-th layer is quantized by finding the closest codebook vector for the current residual, as shown below.
\begin{equation}
    c_l^{t} = \arg\min_k \|\mathbf{r}_l^t - \mathbf{e}_{l,k}^t\|_2, c_l^{v} = \arg\min_k \|\mathbf{r}_l^v - \mathbf{e}_{l,k}^v\|_2,
\end{equation}

\begin{equation}
    \mathbf{r}_{l+1}^t = \mathbf{r}_{l}^t - \mathbf{e}_{l, c_k^t}^t, \mathbf{r}_{l+1}^v = \mathbf{r}_l^v - \mathbf{e}^v_{l, c_k^v},
\end{equation}
where $c_l^{t}$ and $ c_l^{v} $ represent the codewords for text and visual information at the $l$-th layer of residual quantization. $\mathbf{r}_{l+1}^t$ and $\mathbf{r}_{l+1}^v$ are the residual vectors for the next layer. 

However, the quantization of text and visual embeddings described above is performed in a fully independent manner, without any interaction between the two modalities. This leads to several drawbacks: (1) Due to the similarity between certain embeddings in the text and visual domains, codebook collapse may occur, resulting in low utilization; (2) The complementary strengths of text and visual information in representing different aspects of the data are not fully leveraged. To address these issues, we introduce cross-modal contrastive learning by leveraging the multimodal pseudo-labels constructed in the previous section. Specifically, we optimize the residual representations in each layer of the codebooks. More concretely, we use visual pseudo-labels to enhance the residual representations of the text modality, and conversely, use textual pseudo-labels to optimize those of the visual modality. The detailed InfoNCE~\cite{oord2018representation} loss for the $l$-th layer are as follows:
\begin{equation}
    \mathcal{L}_\mathrm{con}^{l, v\rightarrow t} = - \frac{1}{B} \sum_{i=1}^B \log \left(
    \frac{
        \exp \left(\langle\mathbf{r}_i^t, \mathbf{r}_{i, pos}^t\rangle/\tau \right)
    }{
        \sum_{j=1}^B \exp \left( \langle\mathbf{r}_i^t, \mathbf{r}_j^t\rangle/\tau \right)
    }\right),
\end{equation}
\begin{equation}
    \mathcal{L}_\mathrm{con}^{l,t\rightarrow v} = - \frac{1}{B} \sum_{i=1}^B \log \left(
    \frac{
        \exp \left(\langle\mathbf{r}_i^v, \mathbf{r}_{i,pos}^v\rangle/\tau \right)
    }{
        \sum_{j=1}^B \exp \left( \langle\mathbf{r}_i^v, \mathbf{r}_j^v\rangle/\tau \right)
    }\right),
\end{equation}
\begin{equation}
    \mathcal{L}_{\text{con}}^l =  \mathcal{L}_{con}^{l, t \rightarrow v} + \mathcal{L}_{con}^{l, v \rightarrow t},
\end{equation}
where $\mathbf{r}_{i, pos}^t$ and $\mathbf{r}_{i, pos}^v$ represent the positive samples for the $i$-th item in the batch that share the same vision pseudo-label $\mathcal{C}_{\mathrm{vision}}$ and text pseudo-label $\mathcal{C}_{\mathrm{text}}$, respectively. Here, $\langle\cdot, \cdot\rangle$ denotes the inner product, $B$ is the batch size, and $\tau$ is the temperature parameter used in the contrastive loss.

\subsubsection{Cross-modal Reconstruction Alignment}
At the same time, given the $L$ layers of codebooks, the quantized representation can be obtained by summing the corresponding codebook vectors of each layer for the item, denoted as $\mathbf{\hat{z}}^{t} = \sum_{l=0}^{L-1} \mathbf{e}^{t}_{l, c^{t}_k}, \mathbf{\hat{z}}^{v} = \sum_{l=0}^{L-1} \mathbf{e}^{v}_{l, c^{v}_k}$. In order to further utilize the quantized representations from different modalities to refine the codebook representations and balance codebook utilization, we introduce an alignment loss based on contrastive learning. This loss encourages bidirectional alignment between the quantized representations of different modalities for the same item, as formulated below:
\begin{equation}
\mathcal{L}_{\text{align}}^{t \rightarrow v} = -\frac{1}{B} \sum_{i=1}^B \log \left(
    \frac{
        \exp\left( \langle \hat{\mathbf{z}}^t_i,\, \hat{\mathbf{z}}^v_i \rangle/\tau \right )
    }{
        \sum_{j=1}^B \exp\left( \langle \hat{\mathbf{z}}^t_i,\, \hat{\mathbf{z}}^v_j \rangle/\tau \right )
    }
\right), 
\end{equation}
\begin{equation}
\mathcal{L}_{\text{align}}^{v \rightarrow t} = -\frac{1}{B} \sum_{i=1}^B \log \left(
    \frac{
        \exp\left( \langle \hat{\mathbf{z}}^v_i,\, \hat{\mathbf{z}}^t_i \rangle/\tau \right )
    }{
        \sum_{j=1}^B \exp\left( \langle \hat{\mathbf{z}}^v_i,\, \hat{\mathbf{z}}^t_j \rangle/\tau \right )
    }
\right), \
\end{equation}
\begin{equation}
\mathcal{L}_{\text{align}} =  \mathcal{L}_{align}^{t \rightarrow v} + \mathcal{L}_{align}^{v \rightarrow t},
\end{equation}
where $\hat{\mathbf{z}}^t_i, \hat{\mathbf{z}}^v_i$ are the quantization embeddings for text and vision modality of the same item. 

Similar to the RQ-VAE architecture, we decode and reconstruct the quantized representations of different modalities separately. The decoded textual and visual embeddings can be represented as $\hat{\mathbf{t}} = \text{T-Decoder}(\hat{\mathbf{z}}^t)$ and $\hat{\mathbf{v}} = \text{V-Decoder}(\mathbf{z}^v)$, respectively. 
The training of RQ-VAE can be achieved by optimizing both the reconstruction loss and the residual quantization loss, as follows:

\begin{equation}
    \mathcal{L}_{\text{recon}}^{t} = \|\mathbf{t} - \hat{\mathbf{t}}\|_2^2, \mathcal{L}_{\text{recon}}^{v} = \|\mathbf{v} - \hat{\mathbf{v}}\|_2^2, 
\end{equation}
\begin{equation}
\mathcal{L}_{\text{rq}}^{m}=\!\sum_{l=0}^{L-1}\!\bigl(\|\text{sg}[\mathbf{r}^m_l]-\mathbf{e}^m_{l,c^m_k}\|_2^2+\alpha\|\mathbf{r}^m_l-\text{sg}[\mathbf{e}^m_{l,c^m_k}]\|_2^2\bigr),
\end{equation}
\begin{equation}
     \mathcal{L}_{\text{RQ-VAE}}  = \mathcal{L}_{\text{recon}}^{t} + \mathcal{L}_{\text{recon}}^{v} + \mathcal{L}_{\text{rq}}^{t} + \mathcal{L}_{\text{rq}}^{v}, 
\end{equation}
where $m$ denotes different modalities. $\text{sg}[\cdot]$ represents the stop-gradient operation. $\alpha$ is a loss coefficient.  The superscripts $t$ and $v$ denote the textual and visual modalities, respectively.

Finally, the overall training objective for learning the semantic identifier, denoted as $\mathcal{L}_{\text{ID}}$, is defined as follows:
\begin{equation}
    \mathcal{L}_{\text{ID}} = \mathcal{L}_{\text{RQ-VAE}} + \lambda_{\text{con}}^l\sum\limits_{l=0}^{L-1}\mathcal{L}^l_{\text{con}} + \lambda_{\text{align}}\mathcal{L}_{\text{align}},
\end{equation}
where $\mathcal{L}_{\text{RQ-VAE}}$ represents the reconstruction and codebook learning losses for both text and visual modalities, and $\lambda_{con}^l$ and $\lambda_{align}$ are trade-off hyperparameters that balance the contribution of the contrastive loss $\mathcal{L}_{\text{con}}$ and the alignment loss $\mathcal{L}_{\text{align}}$, respectively.

For cases where conflicts occur among certain item IDs, we adopt the same conflict resolution strategy as proposed by \citeauthor{mql}, where codewords are reassigned according to the distance between items and the codebook.

\subsection{Generative Recommendation with Multi-aspect Alignment}
Through the aforementioned trained RQ-VAE model, we obtain discrete semantic IDs for both texts and images, which can be represented as ``\textless a\_1\textgreater\textless b\_2\textgreater\textless c\_3\textgreater'' for text and ``\textless A\_1\textgreater\textless B\_2\textgreater\textless C\_3\textgreater'' for images. For the training of GR models, it is necessary to construct Seq2Seq training data based on these discrete semantic IDs of items. The recommendation model is then trained in a next-token prediction manner. To further optimize the sharing and interaction of information across different modalities, we design both implicit alignment and explicit alignment mechanisms.

\subsubsection{Implicit Alignment for Cross-modal Semantic IDs}
First, we aim for the model to better recognize the commonality between the semantic IDs of different modalities belonging to the same item. To achieve this goal, we align them at the latent space level after encoding. Specifically, based on the encoder-decoder architecture as the GR model, we encode both the textual and visual semantic IDs into latent representations using the encoder of GR model. We then align the representations of different modalities for the same item in the latent space through contrastive learning. Suppose the textual semantic ID of an item is denoted as $t\text{-}sid$ and the visual semantic ID as $v\text{-}sid$. The implementation is as follows,
\begin{equation}
    \mathbf{e}^t = \mathrm{MeanPool(}\text{T5-Encoder}(t\text{-}sid)),
\end{equation}
\begin{equation}
    \mathbf{e}^v = \mathrm{MeanPool(}\text{T5-Encoder}(v\text{-}sid)),
\end{equation}
\begin{equation}
\mathcal{L}_{\text{implicit}}^{t\rightarrow v} = -\frac{1}{B} \sum_{i=1}^B \log \left(
    \frac{
        \exp\left( \langle \mathbf{e}^t_i,\, \mathbf{e}^v_i \rangle/\tau \right )
    }{
        \sum_{j=1}^B \exp\left( \langle \mathbf{e}^t_i,\, \mathbf{e}
        ^v_j \rangle/\tau \right )
    }
\right), 
\end{equation}
\begin{equation}
\mathcal{L}_{\text{implicit}}^{v\rightarrow t} = -\frac{1}{B} \sum_{i=1}^B \log \left(
    \frac{
        \exp\left( \langle \mathbf{e}^v_i,\, \mathbf{e}^t_i \rangle/\tau \right )
    }{
        \sum_{j=1}^B \exp\left( \langle \mathbf{e}^v_i,\, \mathbf{e}
        ^t_j \rangle/\tau \right )
    }
\right),
\end{equation}
\begin{equation}
\mathcal{L}_{\text{implicit}} =  \mathcal{L}_{\text{implicit}}^{t \rightarrow v} + \mathcal{L}_{\text{implicit}}^{v \rightarrow t}.
\end{equation}
\subsubsection{Explicit Alignment with Different Generation Tasks}
Furthermore, for most of the generative models, we can align the representations of images and items by designing different training tasks. Inspired by \citeauthor{mql}, we propose both item-level and sequence-level cross-modal alignment strategies. For item-level alignment, we use the textual semantic ID of an item as input to generate its visual semantic ID, and vice versa, using the visual semantic ID as input to generate the textual semantic ID. For sequence-level alignment, we construct a prediction task where a historical sequence of textual semantic IDs is used to predict the visual semantic ID of the next recommended item. Similarly, a sequence of visual semantic IDs is used to predict the textual semantic ID of the next item. These additional explicit alignment tasks are incorporated into the sequential recommendation training.

\subsubsection{Training Objections and Inference}
For multimodal GR, there are two main tasks for recommendation. The first is to predict the textual semantic ID of the next item based on the historical sequence of item textual semantic IDs. The second is to predict the visual semantic ID of the next item using the historical sequence of item visual semantic IDs. By integrating the aforementioned alignment strategies, the final training objective is formulated as follows:

\begin{equation}
    \mathcal{L}_{\text{rec}} = -\sum\limits_{t=1}^{|y|}logP_{\theta}(y_t|y<t, x) + \lambda_{\text{implicit}}\mathcal{L}_{\text{implicit}}.
\end{equation}

During the inference stage, we generate multiple candidate semantic IDs for different modalities using constrained beam search~\cite{tiger}. Finally, the ensemble of the results~\cite{mql} from two modalities is performed by averaging the scores of both modalities to obtain the final inference result.


\section{Experiment}
We analyze our model's effectiveness and address: 1) \textbf{RQ1}:  How does MACRec compare to state-of-the-art baselines? 2) \textbf{RQ2}: How do different modules impact MACRec's performance? 3) \textbf{RQ3}: What is the impact of our method on the item collision rate? 4) \textbf{RQ4}: What is the impact of our method on the distribution of code allocation? 5) \textbf{RQ5}: How do different hyperparameters affect MACRec? 
\subsection{Experimental Setup}

\noindent \textbf{Datasets. } 
We employ three real-world recommendation datasets, all constructed from the Amazon Product Reviews dataset, which contains user reviews and item metadata collected between May 1996 and October 2018. Specifically, we conducted experiments on datasets from three different categories: ``Musical Instruments'', ``Arts, Crafts and Sewing'', and ``Video Games''. The detailed statistics of these datasets are summarized in Table~\ref{dataset}.
\begin{table}[h]
\small
\centering
\setlength\tabcolsep{2pt}
\begin{tabular}{lccccc}
\toprule
\textbf{Datasets} & \textbf{\#Users} & \textbf{\#Items} & \textbf{\#Interactions} & \textbf{Sparsity} & \textbf{Avg. len} \\
\midrule
Instruments & 17112 & 6250  & 136226  & 99.87\% & 7.96 \\
Arts       & 22171 & 9416  & 174079  & 99.92\% & 7.85 \\
Games      & 42259 & 13839 & 373514  & 99.94\% & 8.84 \\
\bottomrule
\end{tabular}
\caption{Statistics of the datasets. \textbf{Avg. len} represents the average length of item sequences.}
\label{dataset}
\end{table}
\begin{table*}[t]
\setlength\tabcolsep{2.5pt}
\centering

\begin{tabular}{l|l|cccccccccccc}
\toprule
Dataset & Metrics & BERT4Rec & SASRec & FDSA & S$^3$-Rec & MISSRec & P5-CID & VIP5 & TIGER & MQL4GRec & MACRec \\
\midrule
\multirow{5}{*}{Instruments} 
& HR@1      & 0.0450 & 0.0318 & 0.0530 & 0.0339 & 0.0723 & 0.0512 & 0.0737 & 0.0754 & \underline{0.0763} & \textbf{0.0819}$^*$  \\
& HR@5      & 0.0856 & 0.0946 & 0.0987 & 0.0937 & \underline{0.1089} & 0.0839 & 0.0892 & 0.1007 & 0.1058 & \textbf{0.1110}$^*$ \\
& HR@10     & 0.1081 & 0.1233 & 0.1249 & 0.1123 & \underline{0.1361} & 0.1119 & 0.1071 & 0.1221 & 0.1291 & \textbf{0.1363} \\
& NDCG@5    & 0.0667 & 0.0654 & 0.0775 & 0.0693 & 0.0797 & 0.0678 & 0.0815 & 0.0882 & 0.0902 & \textbf{0.0965}$^*$ \\
& NDCG@10   & 0.0739 & 0.0746 & 0.0859 & 0.0743 & 0.0880 & 0.0704 & 0.0872 & 0.0950 & 0.0997 & \textbf{0.1046}$^*$ \\
\midrule
\multirow{5}{*}{Arts}
& HR@1      & 0.0289 & 0.0212 & 0.0380 & 0.0172 & 0.0479 & 0.0421 & 0.0474 & 0.0532 & \underline{0.0626} & \textbf{0.0685}$^*$ \\
& HR@5      & 0.0697 & 0.0951 & 0.0832 & 0.0739 & 0.1021 & 0.0713 & 0.0704 & 0.0894 & \underline{0.0997} & \textbf{0.1046}$^*$ \\
& HR@10     & 0.0922 & 0.1250 & 0.1190 & 0.1030 & \underline{0.1321} & 0.0994 & 0.0959 & 0.1167 & 0.1254 & \textbf{0.1329}$^*$ \\
& NDCG@5    & 0.0502 & 0.0610 & 0.0583 & 0.0511 & 0.0699 & 0.0607 & 0.0586 & 0.0718 & \underline{0.0816} & \textbf{0.0868}$^*$ \\
& NDCG@10   & 0.0575 & 0.0706 & 0.0695 & 0.0630 & 0.0815 & 0.0662 & 0.0635 & 0.0806 & \underline{0.0898} & \textbf{0.0953}$^*$ \\
\midrule
\multirow{5}{*}{Games}
& HR@1      & 0.0115 & 0.0069 & 0.0163 & 0.0136 & \underline{0.0201} & 0.0169 & 0.0173 & 0.0166 & {0.0200} & \textbf{0.0208}$^*$ \\
& HR@5      & 0.0426 & 0.0587 & 0.0614 & 0.0527 & \textbf{0.0674} & 0.0532 & 0.0480 & 0.0523 & 0.0645 & \underline{0.0671} \\
& HR@10     & 0.0725 & 0.0985 & 0.0988 & 0.0903 & \underline{0.1048} & 0.0824 & 0.0758 & 0.0857 & 0.1007 & \textbf{0.1078}$^*$  \\
& NDCG@5    & 0.0270 & 0.0333 & 0.0389 & 0.0351 & 0.0385 & 0.0331 & 0.0328 & 0.0345 & \underline{0.0421} & \textbf{0.0435}$^*$ \\
& NDCG@10   & 0.0366 & 0.0461 & 0.0509 & 0.0468 & 0.0499 & 0.0454 & 0.0418 & 0.0453 & \underline{0.0538} & \textbf{0.0565}$^*$ \\
\bottomrule
\end{tabular}
\caption{Performance comparison of methods on three datasets. The best and second-best results are in bold and underlined, respectively. * denotes statistical significance ($p$-value~$<$~0.05) against the best baseline.}
\label{tab:main}
\end{table*}

\noindent \textbf{Baselines. }
To evaluate our approach, we compare it with representative recent methods, including BERT4Rec~\cite{sun2019bert4rec}, SASRec~\cite{sasrec}, FDSA~\cite{fdsa}, S$^3$-Rec~\cite{zhou2020s3}, MISSRec~\cite{wang2023missrec}, P5~\cite{p5}, VIP5~\cite{vip5}, TIGER~\cite{tiger}, and MQL4GRec~\cite{mql}.

\noindent \textbf{Metrics. }
To assess recommendation effectiveness, we adopt top-k hit rate (HR@K) and normalized Discounted Cumulative Gain (NDCG@K), where K is set to 1, 5, and 10. Consistent with prior studies~\cite{p5,hua2023index}, we utilize a leave-one-out evaluation protocol. Rather than sampling, we conduct full ranking assessments across the entire item collection. 

\noindent \textbf{Implementation Details. }
For the multimodal generative baseline model MQL4GRec, we did not utilize pre-training on millions of additional-category datasets to ensure a fair comparison. Text and image features are obtained using LLaMA and ViT-L/14, respectively. For RQ-VAE, the codebook size $M$ is 256 with 4 levels. We adopt the AdamW optimizer (batch size 1024, learning rate 0.001). The number of clusters $K$ is 512. 
Following \citet{tiger,mql}, we use T5 as the backbone, whose encoder and decoder each have 4 transformer layers with 6 attention heads (dimension 64). The layer-wise contrastive weight $\lambda_{con}^l$ is applied from the third layer onward: $\lambda_{con}^{0,1}=0$, $\lambda_{con}^{2,3}=0.1$. We set $\lambda_{align}=0.001$, $\lambda_{implicit}=0.01$, and temperature $\tau=0.1$. Results are averaged over five random seeds.

\subsection{Performance Analysis (RQ1)}

Table~\ref{tab:main} presents the experimental results of MACRec on three datasets. From the table, we can draw the following conclusions: (1) MACRec achieves the best performance across all three datasets, demonstrating the effectiveness of our proposed approach; (2) MACRec significantly outperforms the state-of-the-art multimodal generative recommendation model MQL4GRec, indicating that the constructed semantic IDs and the cross-modal alignment training strategy can effectively enhance the recommendation performance; (3) Compared with traditional multimodal sequential recommendation models, our model achieves a remarkable improvement in NDCG, suggesting that our multimodal generative recommendation framework can more accurately recommend items that users are interested in.

\begin{table}[htbp]
\centering
\setlength\tabcolsep{5.5pt}
\begin{tabular}{l|ccc}
\toprule
Model & Instruments & Arts & Games \\
\midrule
MACRec & \textbf{0.1363} & \textbf{0.1329} & \textbf{0.1078} \\
w/o $\mathcal{L}_{\text{con}}^l$ & 0.1289 & 0.1283 & 0.1018 \\
w/o $\mathcal{L}_{\text{align}}$ & 0.1310 & 0.1301 & 0.1026 \\
w/o $\mathcal{L}_{\text{implicit}}$ & 0.1312 & 0.1296 & 0.1042 \\
w/o Explicit Alignment & 0.1296 & 0.1299 & 0.1037 \\
\bottomrule
\end{tabular}
\caption{Ablation study (HR@10) on three datasets.}
\label{tab:ablation}
\end{table}
\begin{figure*}[t]
    \centering
    \subfloat[codebook size $M$]{%
        \includegraphics[width=0.16\textwidth]{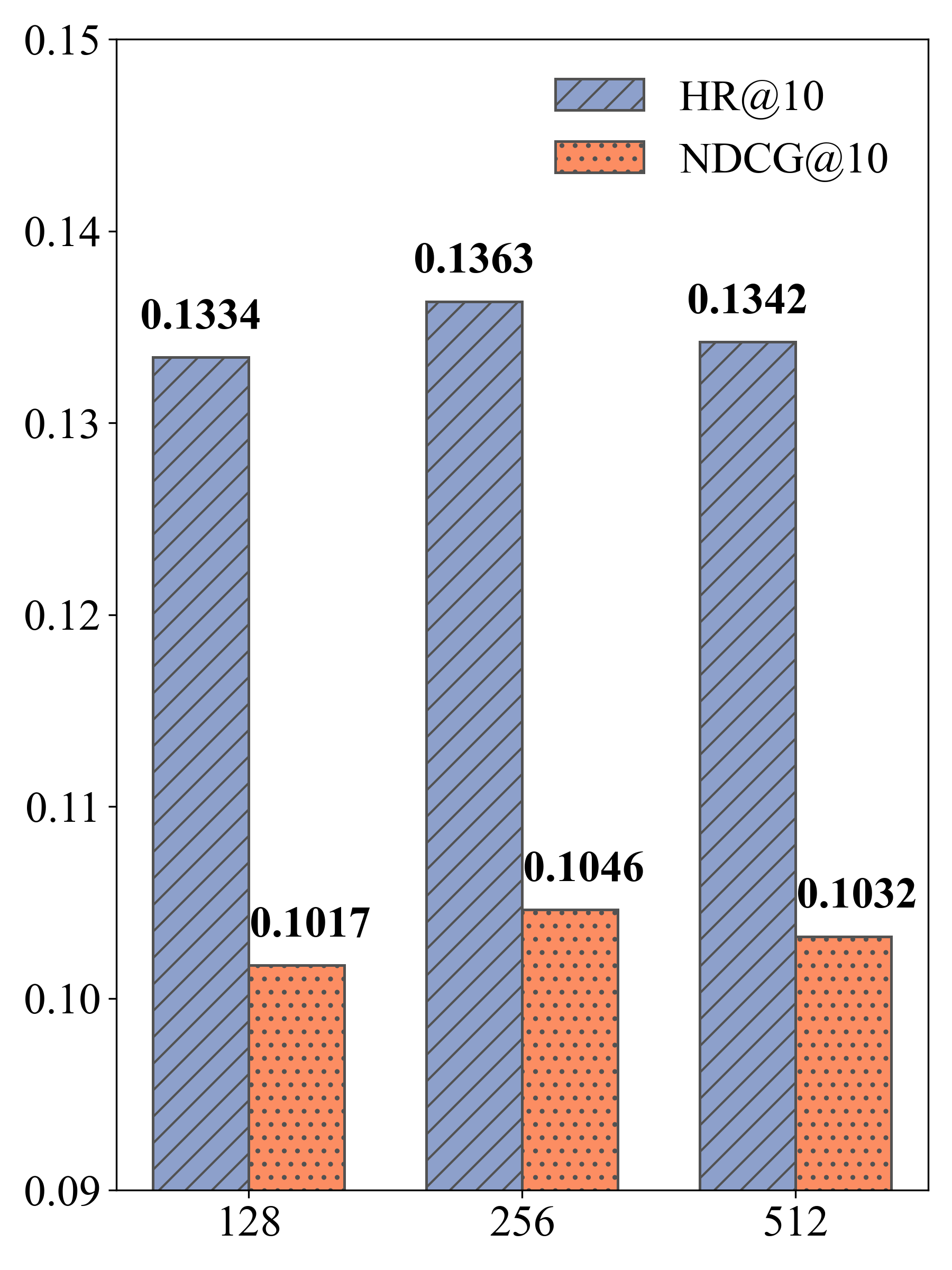}
    }
    \subfloat[semantic ID length]{%
        \includegraphics[width=0.16\textwidth]{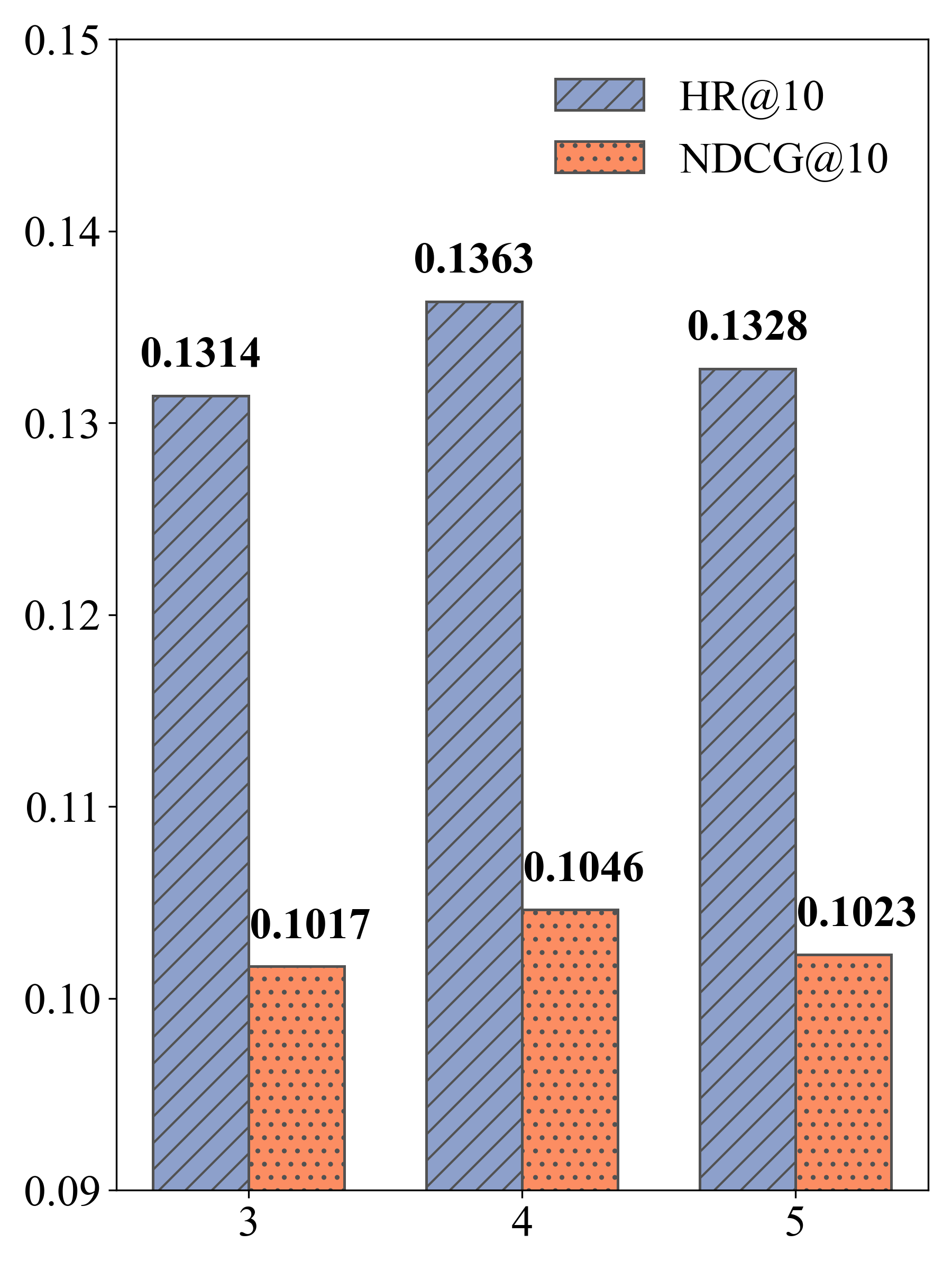}
    }
    \subfloat[start layer of $\mathcal{L}_{con}$]{%
        \includegraphics[width=0.16\textwidth]{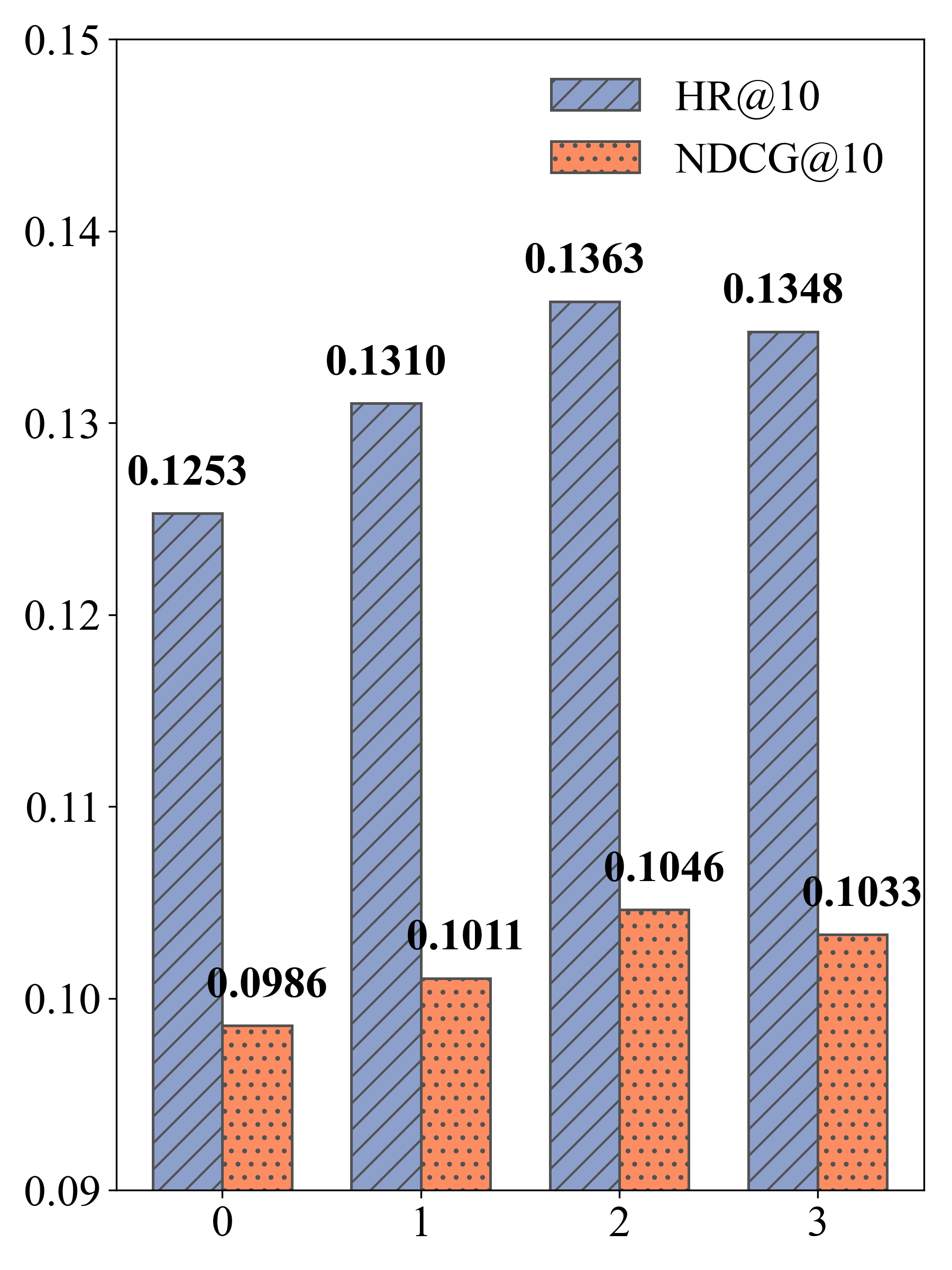}
    }
    \subfloat[$\mathcal{L}_{con}$]{%
        \includegraphics[width=0.16\textwidth]{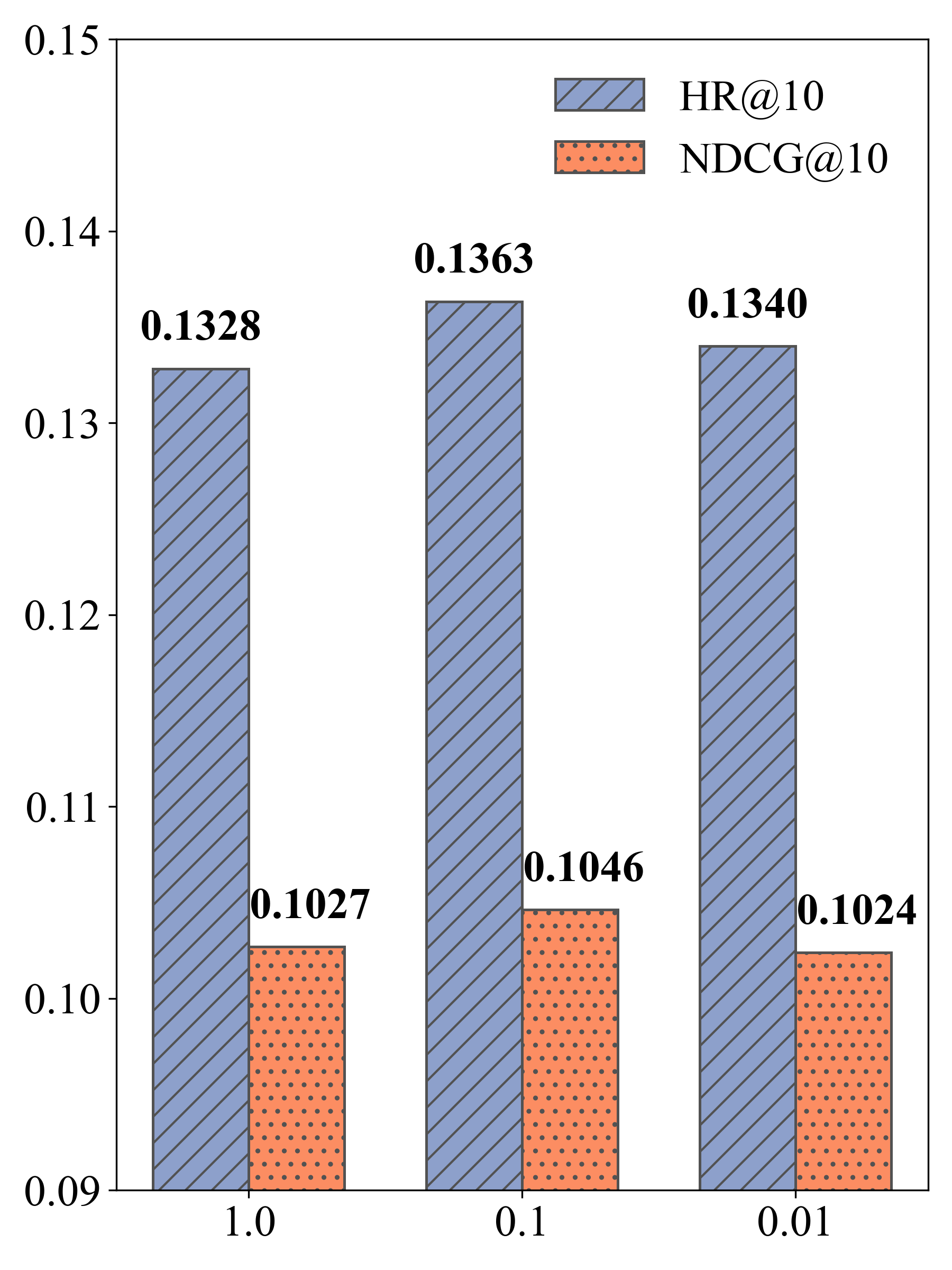}
    }
    \subfloat[$\mathcal{L}_{align}$]{%
        \includegraphics[width=0.16\textwidth]{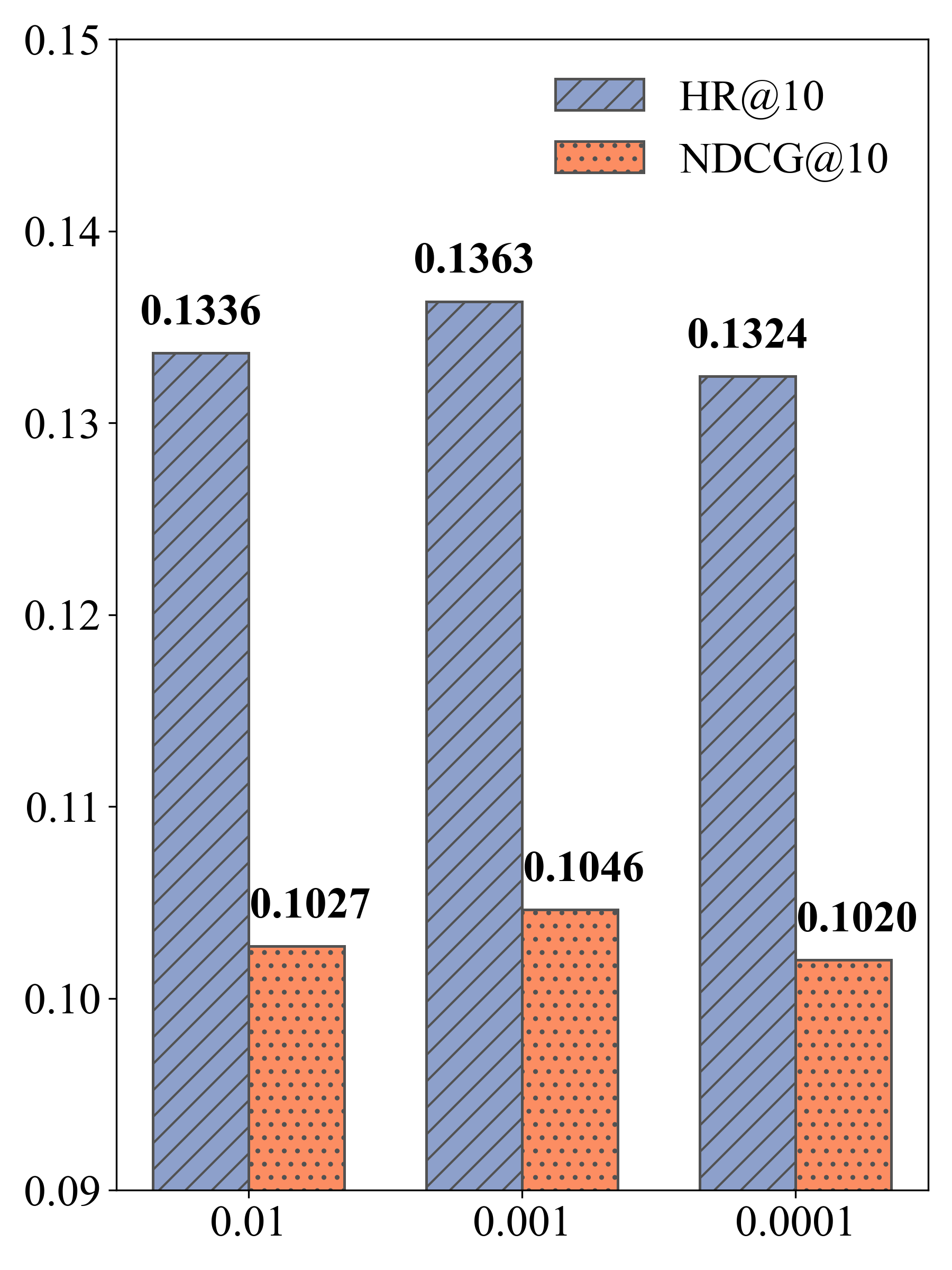}
    }
    \subfloat[$\mathcal{L}_{implicit}$]{%
        \includegraphics[width=0.16\textwidth]{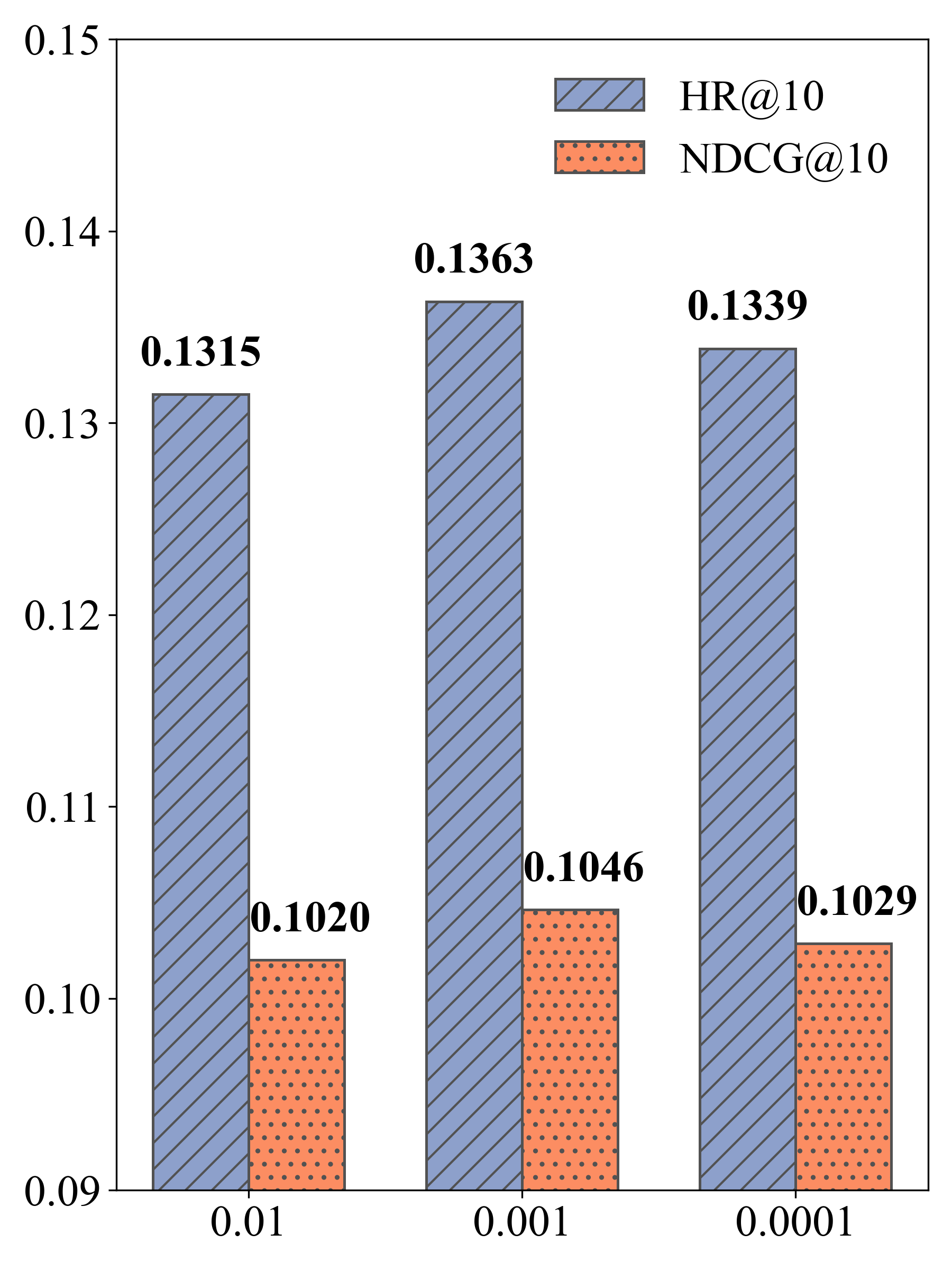}
    }
    \caption{Performance of MACRec over different hyper-parameters on Instruments.}
    \label{fig:params}
\end{figure*}
\begin{figure}[htpb]
    \centering
    \subfloat[MQL4GRec\label{code_left}]{
        \includegraphics[width=0.45\linewidth]{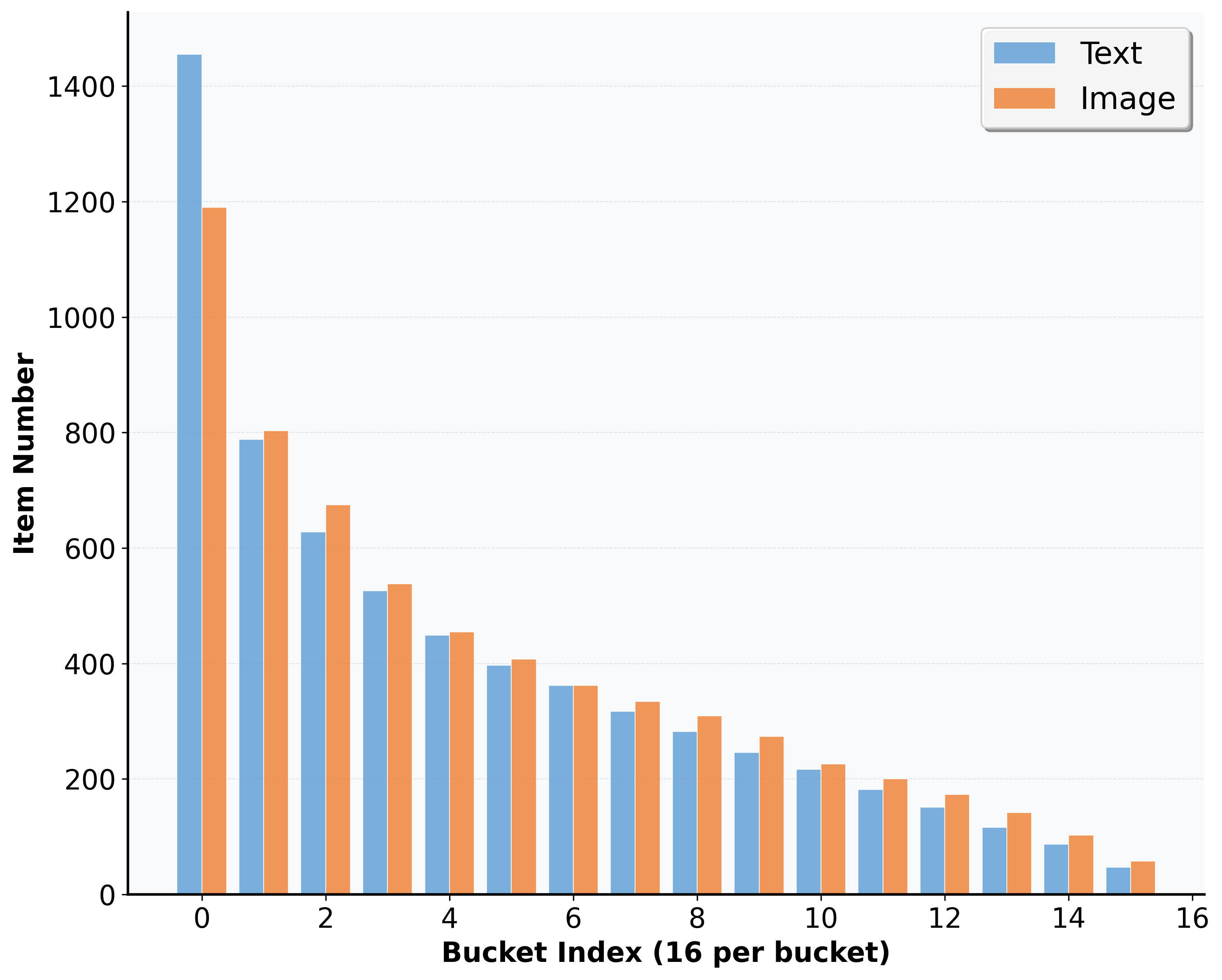}
    }
    \hfill
    \subfloat[MACRec\label{code_right}]{
        \includegraphics[width=0.45\linewidth]{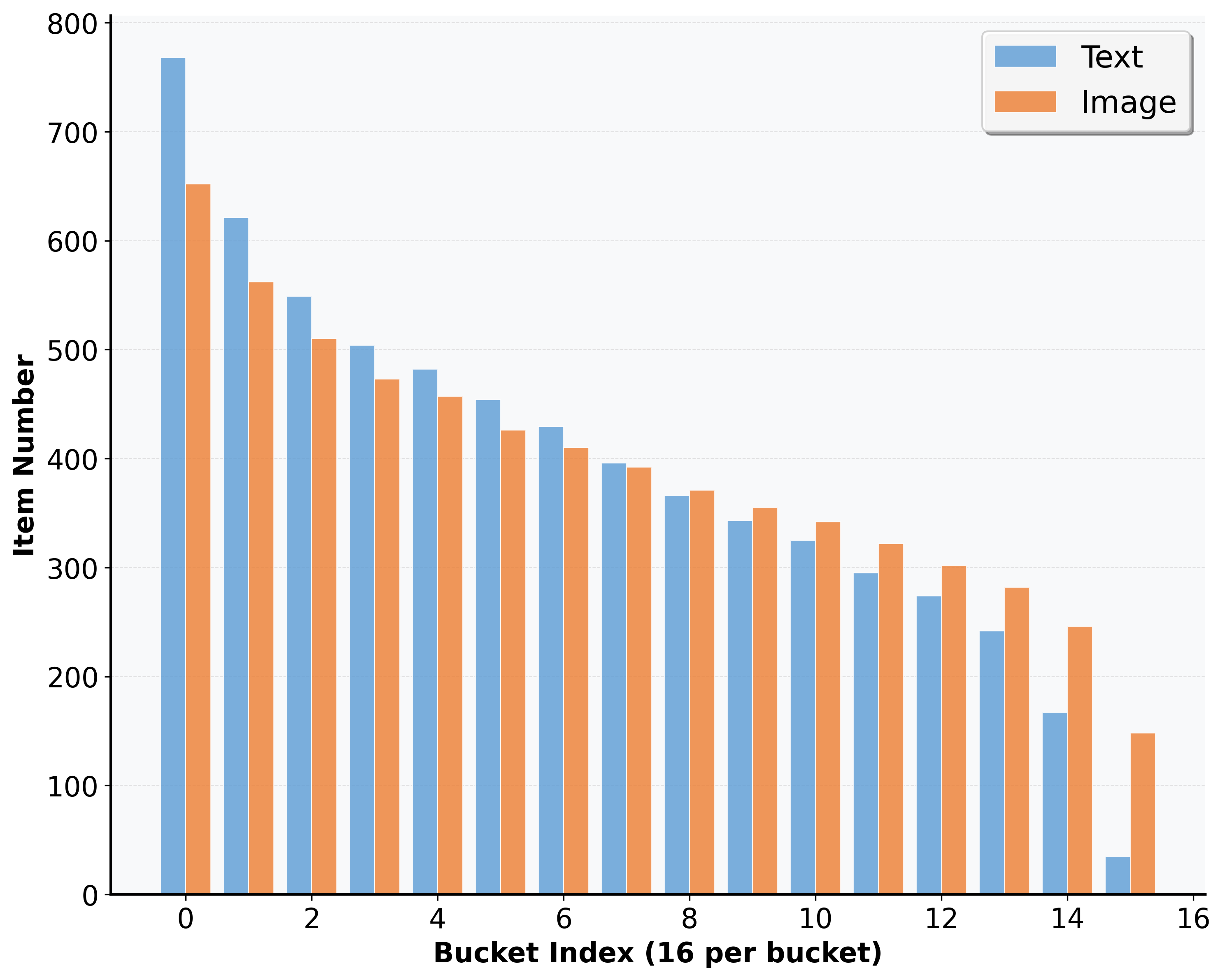}
    }
    \caption{Code assignment distribution on the 2-th RQ layer.}
    \label{fig:code_assignment}
\end{figure}

\subsection{Ablation Study (RQ2)}
Table~\ref{tab:ablation} presents the ablation study results for different loss alignment strategies. From the table, we observe the following: (1) Removing any of the proposed modules leads to a certain degree of performance degradation in recommendation, which demonstrates the effectiveness of our proposed components. (2) Excluding $\mathcal{L}_\mathrm{con}$ results in the largest performance drop across all three datasets, highlighting the effectiveness of our contrastive learning-based cross-modal quantization approach. (3) Both implicit and explicit alignment play important roles in the training of generative models.

\subsection{Item Collision Analysis (RQ3)}
Table~\ref{tab:collision} compares the item collision rates during the quantization process between MACRec and MQL4GRec. The results show that our model can effectively reduce the item collision rate in the quantization process by fully leveraging the complementarity between different modalities. The decrease in collision rate further demonstrates that our method enables a more balanced distribution of items in the codebook, thereby reducing the probability that similar items are encoded with the same semantic ID.

\begin{table}[h]
\centering
\setlength\tabcolsep{1.5pt}
\begin{tabular}{l|cc|cc}
\toprule
\multirow{3}{*}{Dataset} & \multicolumn{2}{c|}{Text} & \multicolumn{2}{c}{Image} \\
\cmidrule(lr){2-3}\cmidrule(lr){4-5}
 & MQL4GRec & MACRec & MQL4GRec & MACRec \\
\midrule
Instruments & 2.76 & \textbf{2.38} & 3.71 & \textbf{3.23} \\
Arts        & 5.15 & \textbf{4.24} & 5.71 & \textbf{5.29} \\
Games       & 3.51 & \textbf{2.91} & 26.10 & \textbf{25.24} \\
\bottomrule
\end{tabular}
\caption{Item ID Collision Rate (\%) comparison between MQL4GRec and MACRec on three datasets.}
\label{tab:collision}
\end{table}

\subsection{Code Assignment Distribution (RQ4)}
Figure~\ref{fig:code_assignment} clearly shows the number of items assigned to each codeword in the second codebook layer for both MACRec and MQL4GRec. The red bars represent the number of items assigned to text semantic IDs, while the blue bars correspond to visual semantic IDs. For clearer visualization, the codewords are sorted in descending order by the number of assigned items, and every 16 codewords are grouped together in a single bar. From the figure, we can observe that MACRec distributes items more evenly across the codewords, indicating a better utilization of codebook capacity and superior semantic representation.

\subsection{Parameter Analysis (RQ5)}

We analyze the effects of key hyperparameters as follows:
1) For codebook size, both very small and very large sizes degrade performance. Small codebooks limit the quantization space and semantic associations, while large ones dilute token exposure, hindering robust representation learning.
2) For semantic ID length, very short IDs fail to capture comprehensive semantics, and overly long IDs complicate learning by expanding the generation space, both reducing performance.
3) When choosing the starting layer for cross-modal contrastive loss $\mathcal{L}_{con}^l$, performance rises then falls; applying it from the first layer erases modality-specific information, while starting from the third layer achieves the best results by letting later VQ layers leverage cross-modal signals to compensate for semantic loss.
4) Each of the other three contrastive losses has an optimal weight: higher weights strengthen modality fusion, while lower weights lead to insufficient cross-modal interaction.


\section{Conclusion}
In this paper, we address the insufficient cross-modal alignment and interaction in current GR methods during semantic ID learning and generative model training. We propose a novel model, \textbf{M}ulti-\textbf{A}spect \textbf{C}ross-modal quantization for generative \textbf{Rec}ommendation (MACRec). MACRec introduces cross-modal interactions at two stages: semantic ID training and generative model training. Specifically, we incorporate cross-modal quantization based on contrastive learning to facilitate the construction of semantic IDs that are both hierarchically semantic and independent. Moreover, we incorporate both implicit and explicit cross-modal alignment in the training process of the generative model, further enhancing the model's understanding of sequential information across different modalities. Extensive experiments demonstrate the superior performance of MACRec in recommendation systems. Finally, we further conduct additional analysis to demonstrate the advantage of our method in codebook utilization.


\section{Acknowledgments}
This work was supported by the National Key Research and Development Program of China under Grant No. 2024YFF0729003, the National Natural Science Foundation of China under Grant Nos. 62176014, 62206266, the Fundamental Research Funds for the Central Universities.

\bibliography{aaai2026}

@article{zhou2025onerec,
  title={OneRec Technical Report},
  author={Zhou, Guorui and Deng, Jiaxin and Zhang, Jinghao and Cai, Kuo and Ren, Lejian and Luo, Qiang and Wang, Qianqian and Hu, Qigen and Huang, Rui and Wang, Shiyao and others},
  journal={arXiv preprint arXiv:2506.13695},
  year={2025}
}

@article{zhou2025onerecv2,
  title={Onerec-v2 technical report},
  author={Zhou, Guorui and Hu, Hengrui and Cheng, Hongtao and Wang, Huanjie and Deng, Jiaxin and Zhang, Jinghao and Cai, Kuo and Ren, Lejian and Ren, Lu and Yu, Liao and others},
  journal={arXiv preprint arXiv:2508.20900},
  year={2025}
}

@article{chen2024fairgap,
  title={FairGap: Fairness-aware recommendation via generating counterfactual graph},
  author={Chen, Wei and Wu, Yiqing and Zhang, Zhao and Zhuang, Fuzhen and He, Zhongshi and Xie, Ruobing and Xia, Feng},
  journal={ACM Transactions on Information Systems},
  volume={42},
  number={4},
  pages={1--25},
  year={2024},
  publisher={ACM New York, NY}
}

@article{chen2025fairdgcl,
  title={FairDgcl: Fairness-aware recommendation with dynamic graph contrastive learning},
  author={Chen, Wei and Yuan, Meng and Zhang, Zhao and Xie, Ruobing and Zhuang, Fuzhen and Wang, Deqing and Liu, Rui},
  journal={IEEE Transactions on Knowledge and Data Engineering},
  year={2025},
  publisher={IEEE}
}

@inproceedings{liu2025multi,
author = {Xiaoyu, Liu and Wu, Yiqing and Han, Ruidong and Zhuang, Fuzhen and Li, Xiang and Lin, Wei},
title = {A Soft-partitioned Semi-supervised Collaborative Transfer Learning Approach for Multi-Domain Recommendation},
year = {2025},
isbn = {9798400720406},
publisher = {Association for Computing Machinery},
booktitle = {Proceedings of the 34th ACM International Conference on Information and Knowledge Management},
pages = {5366–5370},
numpages = {5},
}

@article{liu2025cat,
  title={CAT-ID $^{2}$: Category-Tree Integrated Document Identifier Learning for Generative Retrieval In E-commerce},
  author={Liu, Xiaoyu and Zhang, Fuwei and Wu, Yiqing and Jia, Xinyu and Xia, Zenghua and Zhuang, Fuzhen and Zhang, Zhao and Jiang, Fei and Lin, Wei},
  journal={arXiv preprint arXiv:2511.01461},
  year={2025}
}

@inproceedings{zhang2022mind,
  title={Mind the gap: Cross-lingual information retrieval with hierarchical knowledge enhancement},
  author={Zhang, Fuwei and Zhang, Zhao and Ao, Xiang and Gao, Dehong and Zhuang, Fuzhen and Wei, Yi and He, Qing},
  booktitle={Proceedings of the AAAI conference on artificial intelligence},
  volume={36},
  number={4},
  pages={4345--4353},
  year={2022}
}

@article{zhang2024temporal,
  title={Temporal knowledge graph reasoning with dynamic memory enhancement},
  author={Zhang, Fuwei and Zhang, Zhao and Zhuang, Fuzhen and Zhao, Yu and Wang, Deqing and Zheng, Hongwei},
  journal={IEEE Transactions on Knowledge and Data Engineering},
  volume={36},
  number={11},
  pages={7115--7128},
  year={2024},
  publisher={IEEE}
}

@inproceedings{zhang2025multi,
  title={Multi-level Relevance Document Identifier Learning for Generative Retrieval},
  author={Zhang, Fuwei and Liu, Xiaoyu and Jia, Xinyu and Zhang, Yingfei and Zhang, Shuai and Li, Xiang and Zhuang, Fuzhen and Lin, Wei and Zhang, Zhao},
  booktitle={Proceedings of the 63rd Annual Meeting of the Association for Computational Linguistics (Volume 1: Long Papers)},
  pages={10066--10080},
  year={2025}
}

@inproceedings{zhang2025hiergr,
  title={HierGR: Hierarchical Semantic Representation Enhancement for Generative Retrieval in Food Delivery Search},
  author={Zhang, Fuwei and Liu, Xiaoyu and Jia, Xinyu and Zhang, Yingfei and Xia, Zenghua and Jiang, Fei and Zhuang, Fuzhen and Lin, Wei and Zhang, Zhao},
  booktitle={Proceedings of the 63rd Annual Meeting of the Association for Computational Linguistics (Volume 6: Industry Track)},
  pages={444--455},
  year={2025}
}

@inproceedings{xi2021modeling,
  title={Modeling the sequential dependence among audience multi-step conversions with multi-task learning in targeted display advertising},
  author={Xi, Dongbo and Chen, Zhen and Yan, Peng and Zhang, Yinger and Zhu, Yongchun and Zhuang, Fuzhen and Chen, Yu},
  booktitle={Proceedings of the 27th ACM SIGKDD Conference on Knowledge Discovery \& Data Mining},
  pages={3745--3755},
  year={2021}
}

@inproceedings{xi2019modelling,
  title={Modelling of bi-directional spatio-temporal dependence and users’ dynamic preferences for missing poi check-in identification},
  author={Xi, Dongbo and Zhuang, Fuzhen and Liu, Yanchi and Gu, Jingjing and Xiong, Hui and He, Qing},
  booktitle={Proceedings of the AAAI conference on artificial intelligence},
  volume={33},
  number={01},
  pages={5458--5465},
  year={2019}
}

@article{xi2020graph,
  title={Graph factorization machines for cross-domain recommendation},
  author={Xi, Dongbo and Zhuang, Fuzhen and Zhu, Yongchun and Zhao, Pengpeng and Zhang, Xiangliang and He, Qing},
  journal={arXiv preprint arXiv:2007.05911},
  year={2020}
}

@inproceedings{du2024enhancing,
  title={Enhancing job recommendation through llm-based generative adversarial networks},
  author={Du, Yingpeng and Luo, Di and Yan, Rui and Wang, Xiaopei and Liu, Hongzhi and Zhu, Hengshu and Song, Yang and Zhang, Jie},
  booktitle={Proceedings of the AAAI conference on artificial intelligence},
  volume={38},
  number={8},
  pages={8363--8371},
  year={2024}
}

@article{zhang2022latent,
  title={Latent structure mining with contrastive modality fusion for multimedia recommendation},
  author={Zhang, Jinghao and Zhu, Yanqiao and Liu, Qiang and Zhang, Mengqi and Wu, Shu and Wang, Liang},
  journal={IEEE Transactions on Knowledge and Data Engineering},
  volume={35},
  number={9},
  pages={9154--9167},
  year={2022},
  publisher={IEEE}
}

@inproceedings{li2024large,
  title={Large Language Models for Generative Recommendation: A Survey and Visionary Discussions},
  author={Li, Lei and Zhang, Yongfeng and Liu, Dugang and Chen, Li},
  booktitle={Joint 30th International Conference on Computational Linguistics and 14th International Conference on Language Resources and Evaluation, LREC-COLING 2024},
  pages={10146--10159},
  year={2024},
  organization={European Language Resources Association (ELRA)}
}

@article{oord2018representation,
  title={Representation learning with contrastive predictive coding},
  author={Oord, Aaron van den and Li, Yazhe and Vinyals, Oriol},
  journal={arXiv preprint arXiv:1807.03748},
  year={2018}
}

@article{touvron2023llama,
  title={Llama 2: Open foundation and fine-tuned chat models},
  author={Touvron, Hugo and Martin, Louis and Stone, Kevin and Albert, Peter and Almahairi, Amjad and Babaei, Yasmine and Bashlykov, Nikolay and Batra, Soumya and Bhargava, Prajjwal and Bhosale, Shruti and others},
  journal={arXiv preprint arXiv:2307.09288},
  year={2023}
}

@inproceedings{vit,
  title={An Image is Worth 16x16 Words: Transformers for Image Recognition at Scale},
  author={Dosovitskiy, Alexey and Beyer, Lucas and Kolesnikov, Alexander and Weissenborn, Dirk and Zhai, Xiaohua and Unterthiner, Thomas and Dehghani, Mostafa and Minderer, Matthias and Heigold, Georg and Gelly, Sylvain and others},
  booktitle={International Conference on Learning Representations},
  year={2020}
}

@inproceedings{image_rqvae,
  title={Autoregressive image generation using residual quantization},
  author={Lee, Doyup and Kim, Chiheon and Kim, Saehoon and Cho, Minsu and Han, Wook-Shin},
  booktitle={Proceedings of the IEEE/CVF conference on computer vision and pattern recognition},
  pages={11523--11532},
  year={2022}
}

@article{bin2025real,
  title={Real-time Indexing for Large-scale Recommendation by Streaming Vector Quantization Retriever},
  author={Bin, Xingyan and Cui, Jianfei and Yan, Wujie and Zhao, Zhichen and Han, Xintian and Yan, Chongyang and Zhang, Feng and Zhou, Xun and Wu, Qi and Liu, Zuotao},
  journal={arXiv preprint arXiv:2501.08695},
  year={2025}
}

@inproceedings{zhu2024interest,
  title={Interest clock: Time perception in real-time streaming recommendation system},
  author={Zhu, Yongchun and Chen, Jingwu and Chen, Ling and Li, Yitan and Zhang, Feng and Liu, Zuotao},
  booktitle={Proceedings of the 47th International ACM SIGIR Conference on Research and Development in Information Retrieval},
  pages={2915--2919},
  year={2024}
}

@inproceedings{zhu2025long,
  title={Long-Term Interest Clock: Fine-Grained Time Perception in Streaming Recommendation System},
  author={Zhu, Yongchun and Jiang, Guanyu and Chen, Jingwu and Zhang, Feng and Wu, Qi and Liu, Zuotao},
  booktitle={Companion Proceedings of the ACM on Web Conference 2025},
  pages={1554--1557},
  year={2025}
}

@inproceedings{davidson2010youtube,
  title={The YouTube video recommendation system},
  author={Davidson, James and Liebald, Benjamin and Liu, Junning and Nandy, Palash and Van Vleet, Taylor and Gargi, Ullas and Gupta, Sujoy and He, Yu and Lambert, Mike and Livingston, Blake and others},
  booktitle={Proceedings of the fourth ACM conference on Recommender systems},
  pages={293--296},
  year={2010}
}

@article{amazonrec1,
  title={Two decades of recommender systems at Amazon. com},
  author={Smith, Brent and Linden, Greg},
  journal={Ieee internet computing},
  volume={21},
  number={3},
  pages={12--18},
  year={2017},
  publisher={Ieee}
}

@inproceedings{taobaorec2,
  title={Behavior sequence transformer for e-commerce recommendation in alibaba},
  author={Chen, Qiwei and Zhao, Huan and Li, Wei and Huang, Pipei and Ou, Wenwu},
  booktitle={Proceedings of the 1st international workshop on deep learning practice for high-dimensional sparse data},
  pages={1--4},
  year={2019}
}

@inproceedings{fdsa,
  title={Feature-level deeper self-attention network for sequential recommendation.},
  author={Zhang, Tingting and Zhao, Pengpeng and Liu, Yanchi and Sheng, Victor S and Xu, Jiajie and Wang, Deqing and Liu, Guanfeng and Zhou, Xiaofang and others},
  booktitle={IJCAI},
  pages={4320--4326},
  year={2019}
}

@inproceedings{narm2017,
  title={Neural attentive session-based recommendation},
  author={Li, Jing and Ren, Pengjie and Chen, Zhumin and Ren, Zhaochun and Lian, Tao and Ma, Jun},
  booktitle={Proceedings of the 2017 ACM on Conference on Information and Knowledge Management},
  pages={1419--1428},
  year={2017}
}

@inproceedings{LCRec,
  title={Adapting large language models by integrating collaborative semantics for recommendation},
  author={Zheng, Bowen and Hou, Yupeng and Lu, Hongyu and Chen, Yu and Zhao, Wayne Xin and Chen, Ming and Wen, Ji-Rong},
  booktitle={2024 IEEE 40th International Conference on Data Engineering (ICDE)},
  pages={1435--1448},
  year={2024},
  organization={IEEE}
}

@inproceedings{zhou2020s3,
  title={S3-rec: Self-supervised learning for sequential recommendation with mutual information maximization},
  author={Zhou, Kun and Wang, Hui and Zhao, Wayne Xin and Zhu, Yutao and Wang, Sirui and Zhang, Fuzheng and Wang, Zhongyuan and Wen, Ji-Rong},
  booktitle={Proceedings of the 29th ACM international conference on information \& knowledge management},
  pages={1893--1902},
  year={2020}
}

@article{gru4rec,
  title={Session-based recommendations with recurrent neural networks},
  author={Hidasi, Bal{\'a}zs and Karatzoglou, Alexandros and Baltrunas, Linas and Tikk, Domonkos},
  journal={arXiv preprint arXiv:1511.06939},
  year={2015}
}

@article{liu2024mmgrec,
  title={Mmgrec: Multimodal generative recommendation with transformer model},
  author={Liu, Han and Wei, Yinwei and Song, Xuemeng and Guan, Weili and Li, Yuan-Fang and Nie, Liqiang},
  journal={arXiv preprint arXiv:2404.16555},
  year={2024}
}

@inproceedings{vip5,
  title={VIP5: Towards Multimodal Foundation Models for Recommendation},
  author={Geng, Shijie and Tan, Juntao and Liu, Shuchang and Fu, Zuohui and Zhang, Yongfeng},
  booktitle={The 2023 Conference on Empirical Methods in Natural Language Processing},
  year={2023}
}

@inproceedings{mql,
  title={Multimodal Quantitative Language for Generative Recommendation},
  author={Zhai, Jianyang and Mai, Zi-Feng and Wang, Chang-Dong and Yang, Feidiao and Zheng, Xiawu and Li, Hui and Tian, Yonghong},
  booktitle={The Thirteenth International Conference on Learning Representations},
  year={2025}
}

@inproceedings{letter,
  title={Learnable item tokenization for generative recommendation},
  author={Wang, Wenjie and Bao, Honghui and Lin, Xinyu and Zhang, Jizhi and Li, Yongqi and Feng, Fuli and Ng, See-Kiong and Chua, Tat-Seng},
  booktitle={Proceedings of the 33rd ACM International Conference on Information and Knowledge Management},
  pages={2400--2409},
  year={2024}
}

@article{tiger,
  title={Recommender systems with generative retrieval},
  author={Rajput, Shashank and Mehta, Nikhil and Singh, Anima and Hulikal Keshavan, Raghunandan and Vu, Trung and Heldt, Lukasz and Hong, Lichan and Tay, Yi and Tran, Vinh and Samost, Jonah and others},
  journal={Advances in Neural Information Processing Systems},
  volume={36},
  pages={10299--10315},
  year={2023}
}

@inproceedings{hua2023index,
  title={How to index item ids for recommendation foundation models},
  author={Hua, Wenyue and Xu, Shuyuan and Ge, Yingqiang and Zhang, Yongfeng},
  booktitle={Proceedings of the Annual International ACM SIGIR Conference on Research and Development in Information Retrieval in the Asia Pacific Region},
  pages={195--204},
  year={2023}
}

@article{cui2022m6rec,
  title={M6-rec: Generative pretrained language models are open-ended recommender systems},
  author={Cui, Zeyu and Ma, Jianxin and Zhou, Chang and Zhou, Jingren and Yang, Hongxia},
  journal={arXiv preprint arXiv:2205.08084},
  year={2022}
}

@inproceedings{dnn,
  title={Deep neural networks for youtube recommendations},
  author={Covington, Paul and Adams, Jay and Sargin, Emre},
  booktitle={Proceedings of the 10th ACM conference on recommender systems},
  pages={191--198},
  year={2016}
}

@inproceedings{sasrec,
  title={Self-attentive sequential recommendation},
  author={Kang, Wang-Cheng and McAuley, Julian},
  booktitle={2018 IEEE international conference on data mining (ICDM)},
  pages={197--206},
  year={2018},
  organization={IEEE}
}

@inproceedings{grcn,
  title={Graph-refined convolutional network for multimedia recommendation with implicit feedback},
  author={Wei, Yinwei and Wang, Xiang and Nie, Liqiang and He, Xiangnan and Chua, Tat-Seng},
  booktitle={Proceedings of the 28th ACM international conference on multimedia},
  pages={3541--3549},
  year={2020}
}

@article{liu2024multimodal,
  title={Multimodal recommender systems: A survey},
  author={Liu, Qidong and Hu, Jiaxi and Xiao, Yutian and Zhao, Xiangyu and Gao, Jingtong and Wang, Wanyu and Li, Qing and Tang, Jiliang},
  journal={ACM Computing Surveys},
  volume={57},
  number={2},
  pages={1--17},
  year={2024},
  publisher={ACM New York, NY}
}

@inproceedings{yi2022multi,
  title={Multi-modal graph contrastive learning for micro-video recommendation},
  author={Yi, Zixuan and Wang, Xi and Ounis, Iadh and Macdonald, Craig},
  booktitle={Proceedings of the 45th international ACM SIGIR conference on research and development in information retrieval},
  pages={1807--1811},
  year={2022}
}

@inproceedings{wang2023missrec,
  title={Missrec: Pre-training and transferring multi-modal interest-aware sequence representation for recommendation},
  author={Wang, Jinpeng and Zeng, Ziyun and Wang, Yunxiao and Wang, Yuting and Lu, Xingyu and Li, Tianxiang and Yuan, Jun and Zhang, Rui and Zheng, Hai-Tao and Xia, Shu-Tao},
  booktitle={Proceedings of the 31st ACM International Conference on Multimedia},
  pages={6548--6557},
  year={2023}
}

@inproceedings{p5,
  title={Recommendation as language processing (rlp): A unified pretrain, personalized prompt \& predict paradigm (p5)},
  author={Geng, Shijie and Liu, Shuchang and Fu, Zuohui and Ge, Yingqiang and Zhang, Yongfeng},
  booktitle={Proceedings of the 16th ACM conference on recommender systems},
  pages={299--315},
  year={2022}
}

@inproceedings{liu2018stamp,
  title={STAMP: short-term attention/memory priority model for session-based recommendation},
  author={Liu, Qiao and Zeng, Yifu and Mokhosi, Refuoe and Zhang, Haibin},
  booktitle={Proceedings of the 24th ACM SIGKDD international conference on knowledge discovery \& data mining},
  pages={1831--1839},
  year={2018}
}

@inproceedings{sun2019bert4rec,
  title={BERT4Rec: Sequential recommendation with bidirectional encoder representations from transformer},
  author={Sun, Fei and Liu, Jun and Wu, Jian and Pei, Changhua and Lin, Xiao and Ou, Wenwu and Jiang, Peng},
  booktitle={Proceedings of the 28th ACM international conference on information and knowledge management},
  pages={1441--1450},
  year={2019}
}

@inproceedings{wei2019mmgcn,
  title={MMGCN: Multi-modal graph convolution network for personalized recommendation of micro-video},
  author={Wei, Yinwei and Wang, Xiang and Nie, Liqiang and He, Xiangnan and Hong, Richang and Chua, Tat-Seng},
  booktitle={Proceedings of the 27th ACM international conference on multimedia},
  pages={1437--1445},
  year={2019}
}

\end{document}